\documentstyle[12pt]{article}
\newcommand{\be}{\begin{equation}}
\newcommand{\ee}{\end{equation}}
\newcommand{\bear}{\begin{eqnarray}}
\newcommand{\ear}{\end{eqnarray}}

\date{}
\begin{document}
\begin{titlepage}
\begin{flushright}
HD--THEP--94--16
\end{flushright}
\quad\\
\vspace{1.8cm}
\begin{center}
{\bf\LARGE THE COSMON MODEL FOR}\\
\medskip
{\bf\LARGE AN ASYMPTOTICALLY VANISHING}\\
\medskip
{\bf\LARGE TIME-DEPENDENT COSMOLOGICAL}\\
\medskip
{\bf\LARGE ``CONSTANT''}\\
\vspace{1cm}
Christof Wetterich\\
\bigskip
Institut  f\"ur Theoretische Physik\\
Universit\"at Heidelberg\\
Philosophenweg 16, D-69120 Heidelberg\\
\vspace{1cm}
{\bf Abstract}
\end{center}
We investigate the coupled system of gravity and a scalar with
exponential
potential. The energy momentum tensor of the scalar field induces a
time-dependent cosmological ``constant''. This adjusts itself
dynamically
to become in the ``late'' universe (including today)
proportional to the energy density
of matter and radiation. Possible consequences for the present
cosmology
are shortly discussed. We also address the question of naturalness of
the
cosmon model.

\end{titlepage}
\newpage
Whenever cosmology encounters potential difficulties in the
description of
the present universe cosmologists revive the discussion about the
cosmological constant \cite{1}. The discrepancy between the critical
energy density
expected from inflationary cosmology and lower dynamical estimates
of this density has been attributed to the cosmological constant
\cite{2}.
The discussion also
pertains to the age of the universe \cite{2} and the formation of
structure \cite{3}. In fact, a cosmological constant $\lambda$ of the
order of today's
critical
energy density in the universe $(\lambda\approx (2\cdot 10^{-3}
eV)^4)$
strongly affects the present universe without altering the successful
predictions of the hot big bang model  at early stages of the
evolution of
the universe.

Despite many attempts \cite{4} we have at present no satisfactory
understanding why $\lambda^{1/4}$ should be much smaller than typical
energy scales of the standard model or even the Planck mass $M_p$.
For a
time-independent cosmological  constant it seems even harder to
explain
why it should be of the order of the present energy density. The
latter
depends on the age of the universe rather than on fundamental
constants.
It looks then not very natural that a constant $\lambda$ should have
a
value which equals the energy density just at a time within the
present
cosmological epoch.
In this work we consider a model where the cosmological ``constant''
varies with time such that the
asymptotic solution for late times is characterized by a constant
ratio
$\lambda(t)/\rho(t)$\cite{7A}, \cite{B}. We discuss consequences for
present cosmology and various alternatives how ``early cosmology''
could
have  made a transition to this type of ``late cosmology''. We also
briefly address the question of naturalness of an asymptotically
vanishing
cosmological ``constant''.

We start from the field equations for a scalar field $\varphi$
coupled to
gravity in a homogenous and isotropic  universe (with $k=0$ and $H$
the
Hubble parameter)
\bear
&&\ddot{\varphi}+3H\dot{\varphi}+\frac{\partial V}{\partial
\varphi}=q^\varphi\label{1}\\
&&\dot{\rho}+3H(\rho+p)+q^\varphi\dot{\varphi}=0\label{2}\\
%% FOLLOWING LINE CANNOT BE BROKEN BEFORE 75 CHAR
&&H^2=\frac{1}{6M^2}\left(\rho+\frac{1}{2}\dot{\varphi}^2+V\right)
\label{3}
\ear
The potential  is assumed to  decrease exponentially for large
$\varphi$
\be\label{4}
V(\varphi)=\hat V\exp\left(-a\frac{\varphi}{M}\right).
\ee
where we define  $M^2=M^2_p/16\pi$ and $a>0$ is a free parameter of
the
model\footnote{For  potentials increasing exponentially with the
field we
can obtain positive $a$ by changing the sign of the field.}.
The constant $\hat V>0$
is arbitrary since it can be changed  by a
 shift in $\varphi$.
The scalar potential acts as a cosmological constant (for details see
later).
Exponential potentials arise very naturally in all models of
unification
with gravity as Kaluza-Klein theories, supergravity theories or
string
theories. In higher dimensional theories the scalar could be
associated
with the volume of ``internal space''. The exponential form of the
potential
reflects here the fact that time derivatives in gravity typically
involve
the logarithm of length scales \cite{EXP1}. Similar general arguments
can
be formulated for the exponential form of the potential for the
dilaton
in string theories (or for some of the $SU(3)\times SU(2)\times U(1)$
singlet moduli fields). In the context of inflation exponential
potentials
have been discussed in ref. \cite{EXP2}.

 We also account for possible couplings of the scalar field to matter
\cite{5}
\be\label{5}
q^{\varphi}=g^{-1/2}\left\langle\frac{\delta
S}{\delta\varphi}\right\rangle_{incoherent}=\beta \frac{\rho}{M}.
\ee
Here $\beta$ may effectively depend on the component dominating the
energy
density (e.g. radiation, baryons, or non-baryonic dark matter). For
example, in a baryon dominated universe eq. (\ref{5}) obtains from a
nucleon-scalar coupling corresponding to the Lagrangian
\be\label{6}
{\cal L}_n=-m^{(0)}_n\exp\left(-\beta_n\frac{\varphi}{M}\right)\bar n
n
\ee
similar to the one obtained in string theories. The last term in eq.
(\ref{2}) reflects the change of the effective nucleon mass induced
by a
change of $\varphi$. The late cosmology for a
string theory where one of the scalar modes remains massless can be
formulated in terms of the parameters $a$ and $\beta$. (There are
possibly different $\beta_i$ for the scalar couplings of different
fields
and a completely flat potential is equivalent to the limit
$a\to\infty$.)
As a byproduct of our general discussion one obtains cosmological
constraints on acceptable string theories once $a$ and $\beta$ are
computed
for a given massless mode.\footnote{For bounds on such a coupling in
the
more general case of a scalar field with  mass see ref. \cite{6}.}

The field equations admit the solution
\bear
H&=&\eta t^{-1}\label{7}\\
\rho&=&\rho_0 t^2_0 t^{-2}\label{8}\\
\varphi&=& \varphi_0+\frac{2M}{a}\ln\frac{t}{t_0}\label{9}
\ear
with
\bear
\dot\varphi&=&\frac{2M}{a} t^{-1}\label{10}\\
V(t)&=&V_0 t_0^2 t^{-2}.\label{11}
\ear
All energy densities $(\rho, V,\frac{1}{2}\dot\varphi^2)$ are
$\sim t^{-2}$ and the energy density of the scalar field decreases
simultaneously with $\rho$.\footnote{Cosmologies
with a decaying scalar field were
proposed first in ref. \cite{7A} and studied in a more general
context
in ref. \cite{A} and \cite{B}. An exponential potential without
coupling
to matter $(\beta=0)$ was considered in ref. \cite{7A}, \cite{B}, and
an
extensive discussion for potentials decreasing with a power of
$\varphi$
can be found in ref. \cite{A},\cite{B}.}.
This solution can be interpreted as the asymptotic solution for
$t\to\infty$ (see later). The coefficient $\eta$ obtains from
(\ref{2})
\be\label{12}
\eta=\frac{2}{n}\left(1-\frac{\beta}{a}\right)
\ee
where we use for the equation of state the convention
\be\label{13}
3(\rho+p)=n\rho.
\ee
A deviation from the expansion law of the standard Friedmann universe
occurs only for $\beta$ different from zero. (We only consider
$\beta/a<1$.)

The time-independent ratios of energy densities follow from (\ref{1})
and
(\ref{3})
\bear
V+\frac{\beta}{a}\rho
&=&\frac{2}{a^2}\frac{(3\eta-1)}{\eta^2}M^2H^2\label{14}\\
V+\rho&=&\left(6-\frac{2}{a^2\eta^2}\right)M^2H^2.\label{15}
\ear
With the definitions
\bear\label{16}
\rho_c&=&6M^2H^2,\quad
\Omega_M=\rho/\rho_c\nonumber\\
\Omega_V&=&V/\rho_c,\quad
\Omega_K=\frac{1}{2}\dot\varphi^2/\rho_c
\ear
one obtains
\bear\label{17}
\Omega_M&=&1-\frac{n-2\beta(a-\beta)}{2(a-\beta)^2}\nonumber\\
\Omega_V&=&\frac{n(6-n)-12\beta(a-\beta)}{12(a-\beta)^2}\nonumber\\
\Omega_K&=&\frac{n^2}{12(a-\beta)^2}\nonumber\\
%% FOLLOWING LINE CANNOT BE BROKEN BEFORE 75 CHAR
\Omega_\varphi&=&\Omega_V+\Omega_K=\frac{n-2\beta(a-\beta)}{2(a-\beta)
^2}=1-
\Omega_M.
\ear
Our solution exists only for positive $\Omega_M$ and $\Omega_V$ and
we
have to require
\bear
&&a(a-\beta)>\frac{n}{2}\label{18}\\
&&\beta(a-\beta)<\frac{n(6-n)}{12}.\label{19}
\ear
For the radiation dominated epoch $(n=4)$ $\beta$ presumably
vanishes.
Then our solution only exists for $a^2>2$ (\ref{18}). Eq. (\ref{19})
requires for $n=3$ (non-relativistic matter)
$\beta(a-\beta)<\frac{3}{4}$.
We emphasize that the scalar field contributes to the energy density
of
the universe even for $\beta=0$. It is also interesting to note that
the
scalar field fulfils in our case the ``cosmon condition'' \cite{7}
for the
relation between the dilatation anomaly $\Theta_\mu^\mu$ and the
trace of
the energy momentum  tensor  $T^\mu_\mu$
\be\label{20}
\Theta^\mu_\mu \sim T^\mu_\mu.
\ee
A scalar field with an exponential potential may therefore be called
a
cosmon.

\renewcommand{\labelenumi}{\roman{enumi}.}
Our scenario has several important consequences for cosmology at the
present epoch $(n=3)$:
\begin{enumerate}
\item For $\beta\not= 0$ the age of the universe is modified
\be\label{21}
t_0=\eta H_0^{-1}=\frac{2}{3} H_0^{-1}\left(1-\frac{\beta}{a}\right)
\ee
\item The scalar field contributes to the energy density today.
In units of the critical energy density this contribution is
given by $\Omega_\varphi$ (\ref{17}).
\item The cosmon mediates a new attractive long range force. The
scalar
mass $m_\varphi$ (inverse of the range) is time-dependent and
proportional
 to the Hubble parameter
\be\label{22}
m^2_\varphi=V''(\varphi)=\frac{a^2}{M^2}
V(\varphi)=\frac{9a^2-12\beta
a^2(a-\beta)}{2(a-\beta)^2}H^2
\ee
For all purposes except cosmology the cosmon is effectively
massless\footnote{This distinguishes the present cosmon model from
the
model of ref. \cite{7}.}.
Its coupling to nucleons $\sim\beta_n/M$ (\ref{6}) is of
gravitational
strength (or weaker). At distances small compared to $H^{-1}$ the
effective
Newtons constant for nucleons has a contribution from the scalar
force
\be\label{23} G_N=\frac{1}{M_p^2}(1+4\beta^2_n)\ee
General relativity  distinguishes between the tensor and
scalar component of the attractive force. For distances much smaller
than
$H^{-1}$ the mass term, or more general the influence of the
potential
$V$,
can be neglected. For the matter-dominated epoch the
coupled system of gravity and small scalar fluctuations
around the cosmological value $\varphi(t)$ (\ref{9}) is
in this range identical to the
standard Brans-Dicke \cite{8} theory with
\be\label{24}\omega=\frac{1}{8\beta^2_n}-\frac{3}{2}\ee
(By an
appropriate Weyl scaling the Brans-Dicke theory appears as a gravity
theory with fixed Newtons constant plus a massless scalar field with
universal coupling to matter \cite{5}.) By this analogy we infer the
bound
(from $\omega>500$)
\be\label{25}
\beta_n<0.016\ee
For $a$ of the order one or larger the modification of the age of the
universe
(\ref{21}) can therefore only be minor and is far smaller than the
observational uncertainties in the foreseeable future. We also note
that a
possible non-universal coupling of the scalar field to nucleons (for
example
proportional to baryon number) must be orders of magnitude smaller
than
the
bound (\ref{25}).

\item The scalar field induces a time-dependent cosmological
 constant \cite{C}, \cite{F}, \cite{9}, \cite{7}, \cite{5},
\cite{7A},
\cite{A}, \cite{B}, \cite{D}, \cite{E}. As
we have seen the ratio $V/\rho$ approaches asymptotically the
constant
$\Omega_V/\Omega_M$ (\ref{17}). The kinetic term of the scalar field
also
contributes to the total energy density. There is  an ambiguity in
the definition of the cosmological constant $\lambda$. One may
either put the emphasis of the fact that the energy density of the
scalar field does not participate in gravitational collapse
on scales much smaller than the horizon \cite{R} and
associate the cosmological constant with $\Omega_\varphi$
(\ref{17}). Alternatively, one may concentrate on the
dynamics of the universe as a whole and take the contribution
of the cosmological constant to the energy momentum tensor
proportional to the metric. Then part of the
potential and kinetic scalar energy may be treated as a new form of
nonrelativistic
or relativistic ``matter''.  If we
adopt for the energy momentum tensor the definition
\begin{eqnarray}\label{26}
T_{00}&=&V+\frac{1}{2}\dot\varphi^2+\rho=
\lambda+\rho_\varphi+\rho\nonumber\\
T_{ij}&=&(V-\frac{1}{2}\dot\varphi^2-p)g_{ij}=(\lambda-p_\varphi-p)g_{
ij}
\nonumber\\
p_\varphi&=&(\frac{1}{3}n-1)\rho_\varphi\end{eqnarray}
one obtains the splitting
\begin{eqnarray}\label{27}
\rho_\varphi&=&\frac{3}{n}\dot\varphi^2\nonumber\\
\lambda&=&V+(\frac{1}{2}-\frac{3}{n})\dot\varphi^2=-\frac{6\beta}{a-\b
eta}
M^2H^2\end{eqnarray}
In this language $\lambda$ vanishes for $\beta=0$ (compare also
(\ref{21})),
whereas for $\beta\not=0$ we find $\dot\lambda\sim H^3$ and recover
one of
the
general solutions for a varying cosmological ``constant''
discussed in ref. \cite{9}.
\end{enumerate}

The solution (\ref{7})-(\ref{9}) is not the most general homogeneous
and isotropic universe consistent with (\ref{1})-(\ref{5}). In order
to
establish that a large class of solutions is attracted for large $t$
towards
this particular solution we investigate first the general behaviour
of
small
deviations from (\ref{7})-(\ref{9})\footnote{We can restrict the
discussion to  isocurvature fluctuations  where $H(t)$ can be
computed
from $\rho(t)$ and $\varphi(t)$ (\ref{3}).  For the  zero curvature
cosmologies investigated here one can always take the radius of the
universe arbitrarily large. It is then easy to see that the adiabatic
fluctuations (which grow as usual) decouple and need not to be
considered
explicitly. This reduces  the stability analysis to a
three-dimensional
linear system instead of the five-dimensional system investigated in
ref.
\cite{B}.}
\begin{eqnarray}\label{28}
H&=&\bar H(t)(1+h(t))\nonumber\\
\rho&=&\bar\rho(t)(1+r(t))\nonumber\\
\varphi&=&\bar\varphi(t)+\delta\varphi(t)
\end{eqnarray}
Here $\bar H,\bar\rho$ and $\bar\varphi$ are given by (\ref{7}),
(\ref{8}), and(\ref{9}), and we linearize in the small quantities
$h,r$
and $\delta\varphi$.
We insert (\ref{3})
\be\label{29}
h=\frac{1}{2}(\Omega_Mr-a\Omega_V\frac{\delta\varphi}{M}+2\Omega_K
\frac{\delta\dot\varphi}{\dot{\bar\varphi}})\ee
into  eqs. (\ref{1}), (\ref{2}) and introduce the variables
\begin{eqnarray}\label{30}
\tau&=&\eta\ln (t/t_0)\nonumber\\
\delta\varphi'&=&\frac{d}{d\tau}\delta\varphi,\quad
r'=\frac{d}{d\tau}r\nonumber\\
y&=&\frac{\delta\varphi}{M},\quad
z=\frac{\delta\varphi'}{M}\end{eqnarray}
This gives a coupled system of three linear first-order differential
equations for $x=(z,y,r)$
\be\label{31}
x'=-Gx\ee
with
\be\label{32}
G=\left(\begin{array}{ccc}
A&B&C\\
-1&0&0\\
D&E&F\end{array}\right)\ee
\begin{eqnarray}\label{33}
A&=&3+3\Omega_K-\frac{1}{\eta},\qquad
B=(6a^2-\frac{3}{2}a\gamma)\Omega_V\nonumber\\
C&=&(\frac{3}{2}\gamma-6\beta)\Omega_M,\qquad
D=\frac{n}{\gamma}\Omega_K+\beta\nonumber\\
E&=&-\frac{n}{2}a\Omega_V,\qquad
F=n+\frac{n}{2}\Omega_M-\frac{2}{\eta}+\beta\gamma\end{eqnarray}
where
\be\label{34}
\gamma=\sqrt{12\Omega_K}=\frac{n}{a-\beta}  \ee
The general solution depends on three initial values $r_0=r(t_0)$,
$y_0=
\delta\varphi(t_0)/M$, $z_0=t_0\delta\dot\varphi(t_0)/\eta M$, namely
\be\label{35}
x(t)=\exp\lbrace-\eta\ln\left(\frac{t}{t_0}\right)G\rbrace x_0\ee
The asymptotic behaviour for large $t$ depends on the eigenvalues
$g_i$ of
the matrix $G$. If $Re(g_i)>0$, an arbitrary small $r$ or
$\delta\varphi$
vanishes asymptotically with a negative power of $t$.
In addition, one finds oscillatory behaviour if $Im g_i\not=0$.
For a large range in $a$ and $\beta$ one has indeed $Re(g_i)>0$ and
the
solution (\ref{7})-(\ref{9}) is reached asymptotically for any small
enough
initial value $x_0$.

In particular, for $\beta=0$ and $n=3$ one finds
\bear\label{36}
A&=&\frac{3}{2}+\frac{9}{4a^2},\quad\quad
B=\frac{9}{2}-\frac{27}{8a^2}\nonumber\\
C&=&\frac{9}{2a}(1-\frac{3}{2a^2}),\quad D=\frac{3}{4a}\nonumber\\
E&=&-\frac{9}{8a},\quad\qquad F=\frac{3}{2}-\frac{9}{4a^2}\ear
and
\be\label{37}
\det G=\frac{27}{4}\left(1-\frac{3}{2a^2}\right)\ee
We observe that the stability requirement $\det G>0$ coincides with
the
condition (\ref{18}). For large values of $a(a\to\infty)$ the matrix
$G$
becomes block diagonal in the first two and the third indices, $C, D,
E\to
0$. Then $r$ decays exponentially with $\tau$ such that
\be\label{38}
\rho(t)=\bar\rho(t)\left(1+\frac{r_0t_0}{t}\right)\ee
and the general solution for the scalar field is described by a
decaying
oscillation
\be\label{39}
\varphi(t)=\bar\varphi(t)+\delta\varphi_0\left(\frac{t_0}{t}\right)^{1
/2}
\cos\left(\sqrt{\frac{7}{4}}\ln\frac{t}{t_0}+\alpha_0\right)\ee
The asymptotic solution $(t\to\infty)$ for $a\to\infty$ corresponds
to
standard cosmology since $\Omega_V,\Omega_K\to0$ (\ref{17}). For
finite values of $a$ within the range of stability of the solution
(\ref{7})
-(\ref{9}) the general solution has qualitatively similar properties
as
eqs. (\ref{38}), (\ref{39}), with oscillatory behaviour now also
present
in the time dependence of $\rho$ (for $a^2>\frac{12}{7}$). The
eigenvalues
of $G$ are $\frac{3}{2}$ and $\frac{3}{4}(1\pm
i\sqrt{7-\frac{12}{a^2}})$
and agree with the analysis of ref. \cite{B}.

On the other hand, it is obvious that for initial values
$\varphi(t_0)$
and $\dot\varphi(t_0)$ such that
$V(t_0)\ll\rho(t_0),\dot\varphi^2(t_0)
\ll\rho_0(t_0)$ the scalar field does not influence the time
evolution of
the universe for a certain period after $t_0$. Within a good
approximation
one obtains for this epoch the
standard Friedmann cosmology
\bear\label{40}
H&=&\frac{2}{n}t^{-1}=\eta t^{-1}\nonumber\\
\rho&=&\rho_c=6M^2H^2\ear
As long as $V(t)$ remains small compared to $\rho(t)$, the solution
for
$\varphi$ is approximately\footnote{For $\beta\not=0$ we have not
given
the
most general solution. It looks similar to the solution for $\beta=0$
at
early times and makes then a transition to the logarithm evolution.}.
\be\label{41}
\varphi(t)=\left\lbrace\begin{array}{lll}
\varphi_0+\frac{6\eta^2}{3\eta-1}\beta M\ln\frac{t}{t_0}&\quad{\rm
for}
\quad&\beta\not=0\\
\varphi_0+\delta\varphi_0\left(\frac{t}{t_0}\right)^{1-3\eta}&\quad{\rm
for}
\quad&\beta=0\end{array}\right.\ee
The contribution of the scalar potential to the total energy density
of
the
universe is then
\be\label{42}
\Omega_V(t)=\left\lbrace\begin{array}{lll}
\frac{V_0}{6M^2\eta^2}t^2\left(\frac{t}{t_0}\right)^{-\frac{6\eta^2
\beta a}{3\eta-1}}&\quad{\rm for}
\quad&\beta\not=0\\
\frac{V_0}{6M^2\eta^2}t^2
\exp\left(-a\frac{\delta\varphi_0}{M}\left(\frac{t}{t_0}\right)^{1-3\eta}
\right)
&\quad{\rm for}
\quad&\beta=0
\end{array}\right.\ee
We observe that $\Omega_V$ increases with a power of $t$ for large
values
of $t$.
After a possible short period where the kinetic energy is damped
rapidly,
the scalar field essentially remains constant. It ``sits there and
waits''
until $\Omega_V$ has reached a value of the order one. Then cosmology
makes
a transition to the behaviour described by the solution
(\ref{7})-(\ref{9}).
We conclude that the solution (\ref{40})-(\ref{41}) is unstable at
late
times
since $\Omega_V$ does not remain small. Instead of making a
transition
to the De Sitter or anti-De Sitter universe  as in the  case of a
constant
cosmological ``constant''
we find in our model a much smoother transition to the ``late
cosmology''  given bei (\ref{7})-(\ref{9}).

There is a characteristic time of transition $t_{tr}$ when the
$\rho$-dominated
universe turns over to ``late cosmology'', where $\Omega_V$ reaches a
constant asymptotic value. This transition time depends strongly on
the
initial value of the cosmological constant as given by the potential
energy
density $V(t_0)=V_0$, i.e.
\be\label{43}
t_{tr}\approx M V_0^{-\frac{1}{2}}\ee
One may have the prejudice that $V_0\sim M^4$ and therefore
$t_{tr}\sim M^{-1}$. In this case our modified cosmology would
describe
the
universe at all times after a possible short initial period as, for
example,
inflation. We have, however, no real knowledge on the initial value
$\varphi(t_0)$. Inflation is most likely driven by a field different
from $\varphi$ in order to assure sufficient heating of the universe.
One
may imagine that the system of fields describing inflationary
cosmology
ends with a value $V_0$ many orders of magnitude smaller than $M^4$.
We therefore should also conceive the possibility
that the universe (in the past-inflationary period) is first
described by
Friedman cosmology. Our modified cosmology becomes then relevant only
at
some unknown critical time $t_{tr}$.

As an alternative to the transition from a $\rho$-dominated universe
to
the cosmology with constant $V/\rho$ we may also look at the case
where
the energy density of the universe is dominated at some early epoch
by
contributions
from the scalar field ($\varphi$-dominated universe). Looking for
solutions
of the field equations (\ref{1})-(\ref{3}) with $\rho=0$
we first investigate the approximation where the damping term
$3H\dot\varphi$
in eq. (\ref{1}) can be neglected\footnote{Notice
that $\rho=p=q_\varphi=0$ is always a solution of the field
equations.}.
In this regime the universe undergoes
an exponential expansion similar to scenarios of the inflationary
universe
\cite{10}:
\bear\label{44}
&&V+\frac{1}{2}\dot\varphi^2=E=const\nonumber\\
&&H^2=\frac{E}{6M^2}\ear
Here $\varphi(t)$ is determined implicitly from $V(t)$ via
\be\label{45}
\frac{\sqrt E-\sqrt{E-V(t)}}{\sqrt
E+\sqrt{E-V(t)}}=C_0\exp(-\sqrt{12}aHt)\ee
with $C_0$ a free integration constant. Such an exponential expansion
lasts
as long as $3H\dot\varphi$ can be neglected compated to
$\ddot\varphi$.
The
approximate solution (\ref{45}) implies the ratio
\be\label{46}\frac{3H\dot\varphi}{\ddot\varphi}=\frac{1}{a}
\left(\frac{3E(E-V)}{V^2}\right)^{1/2}\ee
which becomes of the order one once $V(t)$ has decreased sufficiently
according to (\ref{45}). The number of $e$-foldings of the length
scale
of the universe during the period of exponential expansion is roughly
given
by $1/(\sqrt{12}a)$.

Another solution for the pure scalar-gravity system $(\rho=0)$ is
given by
\bear\label{47}
\varphi&=&\varphi_0+\frac{2M}{a}\ln\frac{t}{t_0}\nonumber\\
H&=&\frac{1}{a^2}t^{-1},\quad
V=\frac{2M^2}{a^2}\left(\frac{3}{a^2}-1\right)
t^{-2}\ear
It exists for $a^2<3$ and is a stable attractor. (Small deviations of
$\varphi$
from this solution die out with linear combinations of
$t^{-\lambda_i},\lambda_1=1,\lambda_2=\frac{3}{a^2}-1$.) As long as
$\rho$
is small compared to $V$ this solution approximately determines the
cosmology.
During this epoch $\rho$ decays as
\bear\label{48}
\rho&=&\rho_0\left(\frac{t}{t_0}\right)^{-\delta}\nonumber\\
\delta&=&\frac{n}{a^2}+2\frac{\beta}{a}\ear
For $a(a-\beta)>\frac{n}{2}$ (cf. (\ref{18})) one finds $\delta<2$
and any
initial nonzero $\rho$ becomes comparable to $V$ at late enough time.
The asymptotic solution is then given by eqs. (\ref{7})-(\ref{9}). In
the
opposite case $a(a-\beta)<\frac{n}{2}$ the energy density $\rho$
decreases
faster than $V$. The universe remains $\varphi$-dominated at late
times.
Even an initially $\rho$-dominated universe (40)(41) will in this
case
turn over to a $\varphi$-dominated universe (47)(48) at the
transition
time $t_{tr}$. We observe that the age of the universe is given by
\be\label{49}
t_0=\frac{1}{a^2}H^{-1}_0\ee
if we live today in a $\varphi$-dominated universe. The universe is
older
than for standard cosmology if $a^2<\frac{3}{2}$. The present lower
bound
on $\Omega_M$ indicates, however, that the universe cannot have been
$\varphi$-dominated a long time before the present epoch if
$a(a-\beta)$
is
significantly smaller than $\frac{3}{2}$. This case therefore
necessitates
additional fine-tuning in the initial condition and is in this
respect
similar to the power-law potentials discussed in ref. \cite{A},
\cite{B}.

Let us next ask what type of constraints must be imposed on the
parameter $a$. These constraints depend strongly on the transition
time $t_{tr}$ when the asymptotic ``late cosmology'' with constant
$V/\rho$ begins. Let us first assume that $t_{tr}$ is before the
characteristic time for nucleosynthesis. The standard nucleosynthesis
scenario is then modified by a different speed of the ``gravitational
clock'',
i.e. the ratio between the Hubble parameter and the temperature. We
express temperature and time in units of the (possibly
time-dependent)
nucleon mass and look at the time evolution of
\be\label{50}
\tilde T=T/m_n,\quad\tilde t=tm_n\ee
The critical quantity (corresponding to $H/T$ in standard cosmology)
is given by
\be\label{51}
X=-\tilde T^{-2}\frac{d\tilde T}{d\tilde t}=\left(1+
\frac{\beta_r}{2a\eta}-\frac{2\beta_n}{a\eta}\right)\left(1-
\frac{2\beta_n}{a}\right)^{-1}\frac{H}{T}\ee
(Here we have included for completeness a possible coupling $\beta_r$
of
the scalar field to radiation. We will concentrate on $\beta_r=0$
where
$\eta=1/2$.) Due to the scalar energy density the ratio $H/T$ is
increased by a factor $\Omega_M^{-1/2}$ (cf. (\ref{17})) as compared
to the radiation-dominated universe. Another factor
$(1+4\beta_n^2)^{-1/2}$
arises from the renormalization of the Planck mass (\ref{23}). If
we require $X^2$ to be modified by less than 10 \% as compared to
standard cosmology (more precise bounds can easily be formulated by
noting the equivalence of $X^2$ with additional or missing neutrino
species) one finds $(\beta_r=0)$
\be\label{52}
%% FOLLOWING LINE CANNOT BE BROKEN BEFORE 75 CHAR
\left|\frac{\left(1-4\frac{\beta_n^2}{a^2}\right)^2a^2}{\left(1-2\frac
{\beta_n^2}{a^2}\right)^2(1+4\beta_n^2)(a^2-2)}-1\right|<0.1\ee
or, for very small $|\beta_n/a|,a^2>22$. In contrast, if $t_{tr}$ is
after
the
end of nucleosynthesis the only constraint on $a$ is given by the
existence
of the asymptotic solution (\ref{18}).

For a scenario where  the scalar energy density
dominates the present universe and $a$ is time-independent, we
conclude
that $t_{tr}$ should be later than nucleosynthesis. This type of
cosmology
is characterized by an initial value $\varphi(t_0)$ (e.g. for $t_0$
at the
end of inflation) in a range where $t_{tr}$ (\ref{43}) comes out
between
the end of nucleosynthesis and today. This does not require a
fine-tuning
since $V(t_0)$ can vary over many orders of magnitude. It
necessitates,
however, a small ratio $V_0/M^4$ which, given our lack of knowledge
on
details of the scalar field dynamics during inflation, waits for an
explanation - somewhat similar to the small initial value of
$\Omega-1$ in
Friedman cosmology. As an alternative, we may consider the case
where the parameter $a$ varies as a function of the
curvature  scalar $R$ or the value
of the scalar field. As we will see below the quantity $a$ is closely
related to the breakdown of dilatation symmetry and there is no
reason
why it should be expected to be exactly constant \cite{7A}. In the
course
of the cosmological evolution a dependence of $a$ on $R, H$ or
$\varphi$
would lead to an effective time dependence of $a$. Interesting
cosmological
scenarios with time-dependent $a$ can be imagined: As a first example
$a$
may depend on $\varphi$ in a way that the ratio $V/\rho$ decreases
for
$n=4$ but increases for $n=3$. An appropriate $\varphi$-dependence of
$a$
is not very difficult to construct since the exact exponential
potential
(\ref{4}) seems to be the boundary case between potentials leading to
an
asymptotic $\varphi$-dominated universe (for example power law
potentials
or exponential potentials with small $a$) and potentials implying a
$\rho$-dominated asymptotic universe (see ref. \cite{7A} for an
example). Consistency between small $\Omega_\varphi$ during
nucleosythesis
and an important $\Omega_\varphi$ in the latest period of the
universe can
be achieved in this way. As a different possibility we mention
modifications
of the field equations induced by a curvature or field dependence of
$a$
which are mild enough not to disturb substantially the cosmological
picture
for constant $a$. In this case our asymptotic solution applies with
the
modification that $\Omega_\varphi$ changes slowly. Imagine that $a$
decreases from a value of about twenty during nucleosynthesis to
three
today\footnote{Also the effective value of $\beta_n$ may change with
time and could have been larger during nucleosynthesis.}. This factor
of
seven can be achieved either by a logarithmic
dependence of $a$ on $R$  or by a power law dependence  with a very
small
power.  It is pointless here to discuss details of all these possible
scenarios. Our main conclusion is that for practical purposes we can
simply omit the constraint from nucleosynthesis (\ref{52}) since it
is based on the assumption  of exact constancy of $a$ during many
orders
of magnitude in the cosmological time. A value of $a$ somewhat above
$\frac{3}{2}$ which makes the proposed late cosmology interesting for
observation is perfectly consistent with the idea that an equilibrium
between matter or radiation and the scalar energy density was
established shortly after the end of inflation and prevailed until
today. The initial value $V(t_0)$ would then be close to its
``natural
value'' $H^2(t_0)M^2$.

To close these arguments and elucidate the connection with the fate
of
dilatation symmetry \cite{7A} we briefly present our model in a
somewhat
different language. By an appropriate Weyl scaling the scalar-gravity
model with exponential potential (\ref{4}) can be obtained from the
action
\cite{5}, \cite{7A}.
\be\label{53}
S=-\int d^4 x g^{1/2}\left\lbrace\frac{1}{12}\chi^2 R-\frac{1}{2}
(z-1)\partial_\mu\chi\partial^\mu\chi+\frac{1}{8}\lambda(\chi)\chi^4+{
\cal
L}
_M\right\rbrace\ee
\be\label{54}
\lambda(\chi)=c\chi^{-A}\ee
with the substitution
\be\label{55}
\varphi=\sqrt{12z}X\ln\frac{\chi}{\sqrt{12}M}\ee
If the ``matter Lagrangian'' ${\cal L}_M$ is dilatation-invariant
(for
example, the nucleon mass term is proportional $\chi$), the effective
coupling
of $\varphi$ to matter vanishes $(\beta=0)$ \cite{5}. A value
$\beta\not=0$ therefore reflects dilatation symmetry breaking in the
matter sector and  $\beta$ may
 naturally be small. For $\beta=0$ the ratios of particle masses to
the
Planck mass are independent of the value of $\chi$ and therefore
constant
in
time \cite{5}. If in addition $A$ vanishes, the action is
scale-invariant.
Stability requires $z\geq0$ and for the boundary value
$z=0$ the gravity-scalar system is conformally invariant. We assume
that
quantum effects lead to a breakdown of classical dilatation symmetry
and
induce
$A\not=0$. The quantities $z,A$ and $\lambda$ can be partially
absorbed
by field-redefinitions and the only observable combination turns out
to be
the parameter $a$ in the exponential potential for $\varphi$
\be\label{56} a=(12z)^{-\frac{1}{2}}A\ee
We note that even for small $|A|$ the parameter $a$ can be large if
$z$
is small enough. This requires no fine-tuning since $z=0$ is
singled out by an enhanced (conformal) symmetry.

The quantity $A$ is related to the scale dependence of the coupling
$\lambda$
which can be cast into the form of an evolution equation with
anomalous
dimension
\be\label{57}
\chi\frac{\partial}{\partial\chi}\lambda=-A\lambda+B\lambda^2+...\ee
(The term $\sim B\lambda^2$ and higher order terms would give
corrections
which
are suppressed by powers of $V(t)/M^4$ for large $t$.)
We emphasize that an evolution equation for $\lambda$ with a
nonvanishing
anomalous dimension would not only lead to a phenomenologically
interesting cosmology but solve the whole cosmological constant
problem
in a natural way! In the language with fixed Newton's constant, a
constant
cosmological ``constant''
appears precisely if $V(\chi)\sim\chi^4$. For any potential
increasing
slower
than $\chi^4$ for large $\chi$ the cosmological constant
vanishes\footnote{This
also holds for $V(\chi)$ increasing faster than $\chi^4$, with
$\varphi$
(\ref{55})
replaced by $-\varphi$.} asymptotically! In this language $\chi$
grows
with
increasing time. Since the Planck mass $(\sim\chi)$ grows
faster than $V^{1/4}$ the observable ratio between the cosmological
constant
and the Planck mass decreases \cite{5}. A natural solution of the
cosmological constant problem therefore arises whenever the
asymptotic
behavior of $V(\chi)$ differs from $\chi^4$! This is independent
(for
$A>0$) of the exact form of $V(\chi)$ for small $\chi$. Additional
dilatation symmetry breaking  mass terms  $\sim m^2\chi^2$ or
constant
terms in the potential would not affect the asymptotically vanishing
cosmological ``constant''.

It seems therefore worthwhile to ask if an evolution equation of the
type
(\ref{57}) makes sense. Unfortunately, the answer is not completely
straightforward since we have to deal with the nonrenormalizable
theory
of gravity coupled to the cosmon. In this context we emphasize that
the
scalar field $\varphi$ may
actually not be a ``fundamental field''. It could also correspond to
a
scalar degree of freedom contained in the metric if the effective
action
for gravity is not the pure Einstein action. Similar possibilities
arise
from other
geometrical objects in generalized gravity theories.  At long
distances
these different options all reduce to an effective field theory for
the
graviton-cosmon system.
It is instructive
to recast the evolution equation in the language with field
independent
Newton's constant where it reads\footnote{This equation is to be
interpreted here as a renormalization group equation in the sense of
Coleman
and Weinberg \cite{CW} and not as a field equation.}
\be\label{58}
\frac{d}{d\varphi}V(\varphi)=-\frac{A}{M}V(\varphi)\ee
Let us look what momenta of the quantum fluctuations contribute if we
change
the background field $\varphi$ infinitesimally.
The second derivative of $V$ acts as an effective infrared cutoff for
the quantum fluctuations of the
scalar field. This cutoff  is changed by a shift in $\varphi$ and the
modes
contributing dominantly\footnote{It is not
completely excluded that the contribution of high momentum modes is
also
affected by a shift of the infrared cutoff. We discard this
possibility
here.} have presumably  momentum squared
of the order $V''$.
For $\varphi$ taking values close to those required by the proposed
late
cosmology, typical momenta are of the order of the Hubble parameter.
The
problem of the cosmological constant
is therefore essentially decoupled from what happens at higher
momentum scales  and only involves
the extreme infrared properties of the theory!
For example, the QCD degrees of freedom typically lead to condensates
and
will influence the behavior of $V(\chi)$. These strong interaction
effects
arise
from quantum fluctuations with momenta of the order of a typical QCD
scale.
They affect
the evolution equation for values of $\chi$ much smaller than those
relevant for late cosmology. Electroweak fluctuations or
Planck-length
fluctuations contribute to the evolution equation for the potential
at
even smaller values of $\chi$. For the issue of an asymptotically
vanishing cosmological constant the relevant question concerns the
role
that quantum effects  with length scales much larger than QCD length
scales or even the  inverse electron mass play for the scale
dependence
of the effective
potential\footnote{For a smooth transition to late cosmology we
assume
that the short-distance fluctuations do not generate local minima of
$V(\varphi)$.}.
The only modes which are
expected to contribute to the flow equation are then the graviton and
the
cosmon. (We neglect here the photon and possible massless neutrinos
since they do not couple directly to $\varphi$.) We emphasize that
the
infrared physics of concern here has strictly nothing to do with
possible
ultraviolet divergences of the theory. A suitable method for studying
the
problem could be the concept of the average action \cite{AA}
generalized
to gravity. This tests directly the infrared physics by variation of
an
effective infrared cutoff or averaging over larger and larger
distances.

Waiting for results of this or another method it remains open if the
present approach can lead to a natural solution of the cosmological
constant problem. In the meanwhile we take the cosmon model as an
interesting
phenomenological approach. It seems motivated well enough to merit
detailed studies of its implications for structure formation
or related topics. Such investigations should lead to bounds on
acceptable values of $a$ for present cosmology, similar to those
already obtained for power law potentials \cite{D}, \cite{E}. In
absence
of a complete theory of the fate of the cosmological ``constant''
observation may give important hints!

\end{document}